\documentclass{aa}  

\newcommand{\nust}{{\it NuSTAR}}
\newcommand{\sxrt}{{\it Swift/{\rm XRT}}}
\newcommand{\nice}{{\it NICER}}
\newcommand{\igr}{{\it INTEGRAL}}
\newcommand{\srg}{{\it SRG}}
\newcommand{\art}{ART-XC}
\newcommand{\ero}{{eROSITA}}
\newcommand{\srga}{{SRGA\,J043520.9+552226}}

\def\lum{erg s$^{-1}$}

\usepackage{graphicx}
\usepackage{txfonts}
\usepackage{hyperref}
\usepackage{xcolor}
\usepackage{academicons}
\usepackage{natbib}
\bibpunct{(}{)}{;}{a}{}{,} 

\newcommand{\orcid}[1]{\href{https://orcid.org/#1}{\textcolor[HTML]{A6CE39}{\aiOrcid}}}

\begin{document}

   \title{Peculiar X-ray transient SRGA\,J043520.9+552226/AT2019wey discovered with {\it SRG}/ART-XC}

   \author{I.A. Mereminskiy\thanks{i.a.mereminskiy@gmail.com}\inst{1} \and A.V. Dodin\inst{2} \and A.A. Lutovinov\inst{1} \and A.N. Semena\inst{1} \and V.A. Arefiev\inst{1} \and K.E. Atapin\inst{2} \and A.A. Belinski\inst{2} \and R.A. Burenin\inst{1} \and M.V. Burlak\inst{2} \and M.V. Eselevich\inst{3} \and A.A. Fedotieva\inst{2} \and M.R. Gilfanov\inst{1,4} \and N.P. Ikonnikova\inst{2} \and R.A. Krivonos\inst{1} \and I.Yu. Lapshov\inst{1} \and A.R. Lyapin\inst{1} \and P.S. Medvedev\inst{1} \and S.V. Molkov\inst{1} \and K.A. Postnov\inst{2} \and M.S. Pshirkov\inst{2} \and S.Yu. Sazonov\inst{1} \and N.I. Shakura\inst{2} \and A.E. Shtykovsky\inst{1} \and R.A. Sunyaev\inst{1,4} \and A.M. Tatarnikov\inst{2} \and A.Yu. Tkachenko\inst{1} \and S.G. Zheltoukhov\inst{2} }

   \institute{Space Research Institute, Russian Academy of Sciences, Profsoyuznaya 84/32, 117997 Moscow, Russia
         \and
             Sternberg Astronomical Institute, Moscow M.V. Lomonosov State University,
             Universitetskij pr., 13, 119992, Moscow, Russia
         \and
             Institute of Solar–Terrestrial Physics, Russian Academy of Sciences, Siberian Branch, P.O.Box 4026, 664033, Irkutsk, Russia
         \and
            Max Planck Institute for Astrophysics, 
            Karl-Schwarzschild-Str. 1,
            Postfach 1317,
            D-85741 Garching, Germany}


    \abstract 
    {During the ongoing all-sky survey, the {\it Mikhail Pavlinsky} ART-XC telescope on board the \srg\ observatory should discover new X-ray sources, many of which  can be transient. Here we report on the discovery and multiwavelength follow-up of a peculiar X-ray source SRGA\,J043520.9+552226  = SRGe J043523.3+552234 -- the high-energy counterpart of the optical transient AT2019wey.}
    {Thanks to its sensitivity and the survey strategy, the {\it Mikhail Pavlinsky} ART-XC telescope uncovers poorly studied weak transient populations. Using a synergy with current public optical surveys, we are aiming at revealing the nature of these transients to study its parent populations. The \srga\ is the first transient detected by \art\, which has a bright optical counterpart suitable for further studies. } 
    {We have used available public X-ray and optical data and observations with \srg, \igr, \nust, \nice\ and ground-based telescopes to investigate the source spectral energy distributions at different phases of the outburst.}
    {Based on X-ray spectral and timing properties derived from space observations, optical spectroscopy and photometry obtained with the 2.5-m and RC600 CMO SAI MSU telescopes, we propose the source to be a black hole in a low-mass close X-ray binary system.} 
    {}

   \keywords{X-rays: binaries -- 
    X-rays: individuals: SRGA\,J043520.9+552226
               }
 \titlerunning{{\it SRG}/ART-XC discovery of SRGA\,J043520.9+552226}
\authorrunning{I. Mereminskiy et al.} 

   \maketitle
%
\section{Introduction}

Since December 2019, the {\it Spectrum-Roentgen-Gamma} observatory (\srg, \citealt{sunyaev21}) has been performing a continuous all-sky survey in X-rays using its two instruments: the \ero\ \citep{2021A&A...647A...1P} and {\it Mikhail Pavlinsky} ART-XC telescopes \citep{2021arXiv210312479P}. This survey is expected to be the deepest one in both soft and hard X-rays. Along with already known X-ray sources, both telescopes detect new objects, including transient ones. Particularly interesting are populations of weak transients, not easily observable by current X-ray all-sky monitors and wide field telescopes (e.g., \igr, {\it Swift}/BAT, {\it MAXI}). One of the best examples of such a population are very faint X-ray transients (VFXTs, see, e.g.  \citealt{muno05, wijnands06}) that reach a peak luminosity of  $10^{34}-10^{36}$ \lum\ during outbursts. VFXTs have been unveiled in the recent years by extensive monitoring programs of the Galactic Centre region. Now, with the advent of \art\, we have an opportunity to search for similar families of dim sources over the entire sky.  

Since its launch, the {\it Mikhail Pavlinsky} \art\ telescope revealed about dozen of outbursts from known objects \citep[see, e.g.][]{mereminskiyIGR17379,mereminskiy4U1755}, as well as discovered several new transients \citep{mereminskiyartgc, semena20,2021ATel14357....1S}. Most of these new sources, although interesting, were difficult to study, given the lack of established optical counterparts, which makes it practically impossible to unveil the nature of these sources. 

On Mar 18, 2020, \srg/ART-XC discovered a bright ($F_{4-12 keV}\approx 10^{-11}$ erg cm$^{-2}$ s$^{-1}$) X-ray source \srga\ at the position coinciding with an orphan (i.e. not associated with any background galaxy) optical transient \href{https://wis-tns.weizmann.ac.il/object/2019wey}{AT2019wey}, which was found earlier in ZTF and ATLAS surveys. The source was also detected by \srg/\ero\ in the soft and standard X-ray bands, which provided more accurate localisation, reliably identifying the new X-ray source with AT2019wey \citep{mereminskiyat2019wey}. This allowed us to initiate follow-up optical observations, which were supported by different ground-based telescopes and other X-ray observatories. Based on these observations \srga\ was established as a new Galactic microquasar in a low-mass X-ray binary system (LMXB) showing a resolved compact jet \citep{yadlapalli20}, unusual relation $L_{\rm X}-L_{\rm opt}$ between X-ray and optical emissions and the strong reflection \citep[see][]{yao20mw,yao20xray}.    

In this paper, in Section~\ref{sec:discovery} we report on the discovery of a new transient \srga; in Section ~\ref{sec:observations} comprehensive X-ray and optical observations of the source are described. The observed properties are summarised in Section~\ref{sec:obsprop}. In Section~\ref{sec:disc} we discuss the features and the nature of the source, and in the last Section~\ref{sec:summary} our conclusions are formulated.

\section{Discovery of \srga\ = SRGe J043523.3+552234/ AT2019wey}
\label{sec:discovery}

The \srga\, was first detected on Mar 18, 2020 by the \art\, near-real time analysis pipeline and identified as a new X-ray transient. The source was also detected by the \ero\ telescope which provided a more accurate position and spectral information in the low-energy part of its X-ray spectrum. An improved position of the source (RA=68.8472, Dec.=55.3760, J2000, the 68\% error circle radius 5\arcsec) was just $\sim0.8$\arcsec off the position of \href{https://wis-tns.weizmann.ac.il/object/2019wey}{AT2019wey} --  the optical transient discovered by ZTF \citep{ztf19} on Dec 2, 2019, and later independently detected by ATLAS \citep{tonry18} and {\it GAIA } \citep{gaia}. This positional coincidence enabled us to identify the discovered X-ray source \srga\ = SRGe J043523.3+552234, as an X-ray counterpart of AT2019wey. The X-ray spectrum turned out to be relatively hard and absorbed (\citet{mereminskiyat2019wey}, see Sec.~\ref{sec:obsprop} for details), indicating that a source is not a nearby one.

To investigate the nature of the source, we initiated quick follow-up observations at the 1.6-m Sayan telescope \citep{2016AstL...42..295B}. 
The first optical spectrum was obtained on Mar 19, 2020. Based on a low signal-to-noise, featureless blue spectrum, \citet{lyapin20} proposed the source to be a blazar in the outburst. However, this conclusion was considered initially questionable due to unstable weather conditions resulted in a poor-quality optical spectrum. 

A high-quality optical spectrum of the source was acquired on the 2.5-m telescope of Caucasian Mountain Observatory of Sternberg Astronomical Institute of Moscow State University (CMO SAI MSU) on Mar 27, 2020. The spectrum exhibited the Balmer hydrogen lines  (H$_{\alpha}$-H$_{\epsilon}$) in absorption at rest implying the Galactic origin of the transient. 

Most bright galactic X-ray transients reside in binary systems in which an increase of the X-ray emission is related to the onset of the accretion onto a compact object (white dwarf, neutron star or black hole). We, therefore, inspected archival optical observations in which secondary star could have been detected. 

Earlier, this sky area was observed by the SDSS \citep{sdssxii} and  Pan-STARRS1 \citep{panstarrs1} surveys but no source was found at the position of AT2019wey. This enables constraining the presence of a hot, massive optical component in this system. Both surveys give an upper limit on the brightness of \srga/AT2019wey in quiescence at $\approx$23 mag in the $g$-filter. The Galactic extinction along the line of sight is $E(g-r) = 0.7, A_{g}=3.5$, \citep{green19}, which enables us to exclude any star with the spectral class earlier than F0V ($M_{g} = 3.1$, the absolute magnitudes for main-sequence stars of different spectral classes were taken from \citealt{bilir09}) up to a distance of about 20 kpc. Therefore, the source can not be a Galactic high-mass X-ray binary (HMXB). 

On the other hand, the typical K5V optical star in a  low-mass X-ray binary \citep{smith98} 
with $M_{g}=8.3$ located within 2.5 kpc from the Sun would be detected. Therefore, the source is unlikely to be a nearby LMXB. Using the $\delta V$-period relation from \citet{shahbaz98} and assuming that $\Delta V\approx\Delta g \gtrsim 6$, one could place an upper limit on a binary period of $\sim 15$ hours. Given above considerations -- the source galactic coordinates ($l\simeq151.16$, $b\simeq5.30$) and the lack of a luminous star in the archival data -- one can assume that \srga\, is a LMXB or a cataclysmic variable (CV) located in the far part of the Perseus spiral arm or in the outer arm.  

The discovery of the new transient \srga\ initiated a chain of follow-up observations from X-rays to the optical and radio. Part of these multiwavelength observations were summarised in \citet{yao20mw}. Those authors placed a tight constrains on the secondary star mass $M_{2} \lesssim 0.8 M_{\odot}$ and source distance of $1~{\rm kpc}\, \lesssim D \lesssim 10~{\rm kpc}$. The X-ray behaviour was found to be typical for black-hole LMXBs undergoing a "hard-only" outburst \citep{watchdog16}, with the source slowly transitioning from a 'low-hard' state to a 'hard-intermediate' state \citep{yao20xray}. Finally, interferometric radio observations discovered a variable, extended source indicating powerful jets in this system \citep{yadlapalli20}. 

The above arguments suggest that the \srga\ = SRGe J043523.3+552234, is another Galactic microquasar in a LMXB system.

\begin{figure*}[!t]
\centering
   \includegraphics[width=17cm]{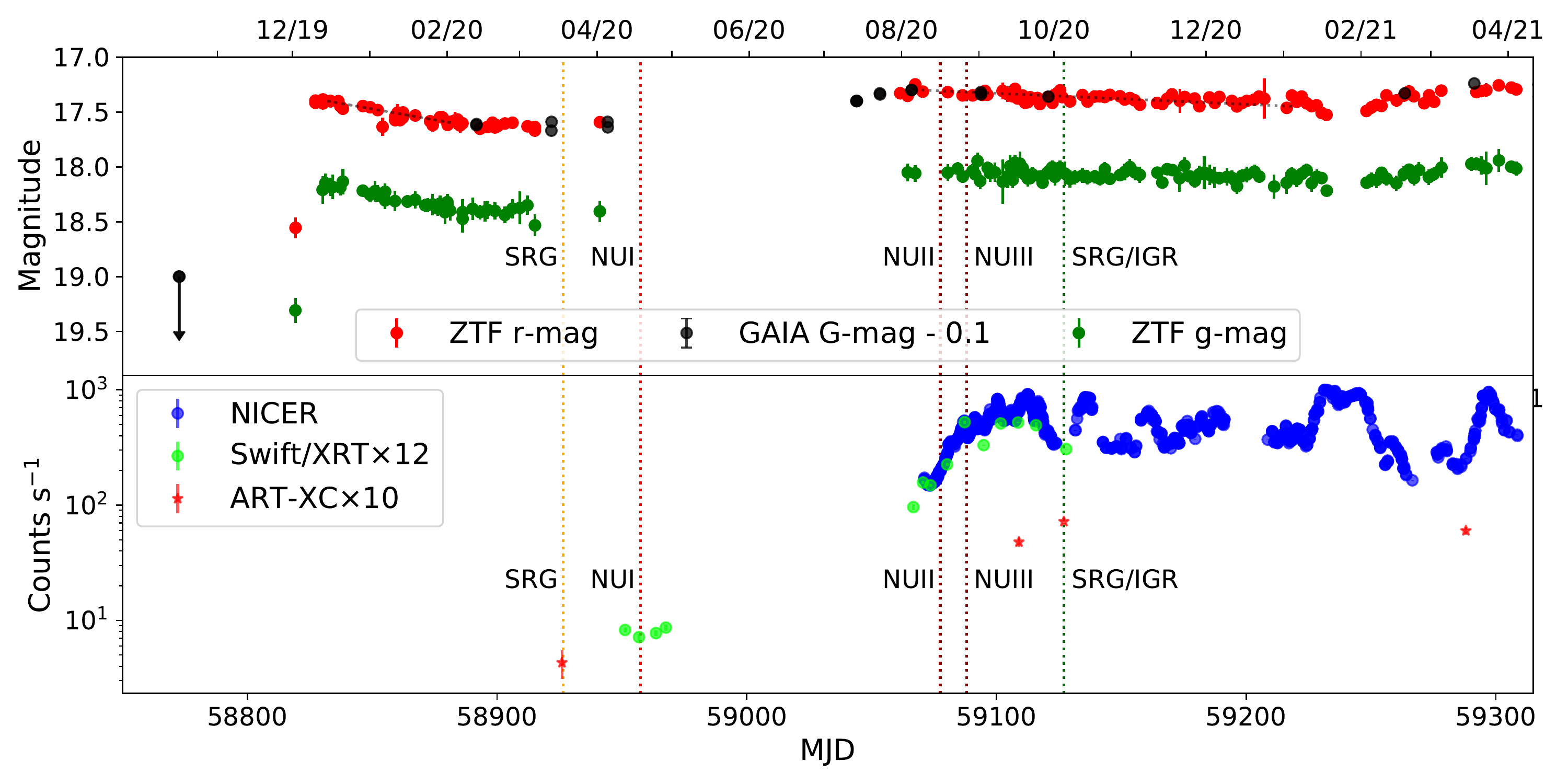}
     \caption{{\it Upper panel}: light curve of AT2019wey as observed by ZTF and {\it Gaia}. The dotted vertical lines denote times when the source was observed by \srg, \nust\, and \igr. The dotted black lines shows a linear decay with 4 mmag/day for winter-spring, 2020 observations and 1 mmag/day for summer-autumn, 2020. Ticks on top of the panel shows beginnings of the corresponding months in MM/YY format.\\
    {\it Lower panel}: \art (red points) , {\it NICER} (blue points) and \sxrt\,(green points) count rates in the broad energy bands (4-12 keV for \art, 0.4-9 keV for {\it NICER} and 0.3-9 keV for \sxrt). The \sxrt\, data are arbitrarily scaled to roughly match the {\it NICER} observations and \art\, data are scaled for clarity. The dotted vertical lines show times of \srg, \nust\ and {\it INTEGRAL} observations.}
     \label{fig:opt_x_lc}
\end{figure*}

\section{Observations and data reduction}
\label{sec:observations}

\subsection{X-ray observations}

\subsubsection{{\it Spectrum-Roentgen-Gamma}}

\srga\xspace has been observed in total four times by the \srg\ observatory, including three observations during the first, second and third all-sky surveys in March, 2020, September, 2020 and March, 2021. These observations were relatively short, so that not much data could be extracted. The  \ero\ raw  data  were  processed  by  the  calibration pipeline based on the \ero\ Science Analysis Software System (eSASS) \citep{brunner18esas}. During the second and third observations the source was bright enough for photon pile-up to become important, therefore we did not use these data, limiting ourselves to only the first scan.

The fourth \srg\ observation was performed on 5-6 Oct, 2020, during the survey break intended for an orbital correction manoeuvre. The \ero\ telescope, with its detectors working at cryogenic temperatures, was turned off during this observation to prevent a possible contamination by the engine exhaust, therefore only \art\ data were available. The observation was performed in the pointing mode with an exposure of $\simeq15$ ks. These observations had been proposed for cross-calibration purposes and were performed simultaneously with the {\it INTEGRAL} ones. 

We used ART-XC pipeline {\sc artproducts} v0.9 with the {\sc caldb} version 20200401 to extract the source spectrum in the 4-30 keV band and a lightcurve in the 4-12 keV energy band from a circular region ($R=135\arcsec$) centred on the source. The spectrum was grouped in order to have at least 30 counts per energy bin, at high energies we used a custom binning to achieve a better signal-to-noise ratio. 

\subsubsection{{\it Swift/XRT}}

Monitoring observations of \srga\, were carried out by \sxrt\, \citep{burrows05} on board the {\it Neil Gehrels Swift} observatory \citep{gehrels04}. Data from \sxrt\, were processed using the online tool \citep{evans09}. The \sxrt\, spectra obtained in the photon-counting mode were used in the 0.3-9 keV range (0.8-9 keV for the windowed mode) and were grouped also to have at least 30 counts per energy bin.

\subsubsection{{\it NICER}}

The \nice\, \citep{gendreau12} telescope, mounted on the {\it International Space Station} obtained a detailed light curve during the outburst brightening. High-level \nice\, products were extracted using the latest {\texttt HEASOFT 6.28} version. To trace the outburst profile, we extracted the overall 0.4-9 keV light curve binned per 1 ks. No background was subtracted, with the source dominating above the mean background by a factor of one hundred or more. 

\subsubsection{{\it NuSTAR}}

Three long observations by {\it NuSTAR} \citep{harrison13} (ObsIDs: 90601315002, 90601315004, 90601315006 hereafter denoted as NUI, NUII and NUIII, respectively) were used to construct broadband spectral energy distributions (SEDs). Spectra and light curves were extracted using \texttt{NuSTARDAS v2.0.0}. Spectra were extracted from circular regions with 72\arcsec, 105\arcsec\, and 75\arcsec\, correspondingly, background regions were selected at the same chip, where possible.

\subsubsection{{\it INTEGRAL}}

We initiated a long Target-of-Opportunity (ToO) observations of \srga\, with the {\it INTEGRAL} observatory \citep{winkler03} simultaneously with the pointed \art\, observations and short (1.5 ks) {\it Swift} ToO on October 4-6, 2020 (rev.2282).
Observations were performed in the hexagonal pointing pattern (HEX) providing a nearly uninterrupted monitoring of the source variability during the entire 180 ks observational run. The 30-150 keV spectrum was extracted from the IBIS/ISGRI \citep{ubertini03, lebrun03} data using a proprietary analysis package developed at Space Research Institute (details available in \citet{krivonos10,2014Natur.512..406C}). We also produced long time series with 100s resolution using IBIS/ISGRI data in the 30-60 keV band and JEM-X \citep{lund03} data in the 3-12 keV band using the {\texttt OSA 11.1} package. 

\subsection{Optical observations}

\subsubsection{{\it 1.6-m Sayan telescope}}

We used the 1.6-m Sayan telescope of the Institute of Solar–Terrestrial Physics (ISTP) RAS observatory to perform initial observations of the source following its discovery. The spectrum with a 7{\AA} resolution in the 4000--9500{\AA} band with a mean signal-to-noise ratio of 20 was obtained with the ADAM spectrometer \citep{afanasiev16, 2016AstL...42..295B}. Unfortunately, the seeing conditions were unfavourable during these observations.  

\subsubsection{{\it CMO SAI MSU}}


\begin{table}
\centering
\caption{Near infrared $J,H,K$ photometry of AT2019wey as observed by ASTRONIRCAM CMO SAI MSU. Magnitudes in Vega system.}
\begin{tabular}{c | c | c | c } 
 \hline
 MJD & J, $mag$ & H, $mag$ & K, $mag$\\ 
 \hline
59088.98 & $15.61\pm{0.04}$ & $15.13\pm{0.02}$ & $14.60\pm{0.05}$\\
59090.03 & $15.58\pm{0.02}$ &                  & $14.64\pm{0.04}$\\
59091.06 & $15.65\pm{0.02}$ & $15.21\pm{0.03}$ & $14.71\pm{0.04}$\\
59093.96 & $15.60\pm{0.01}$ & $15.15\pm{0.04}$ & $14.64\pm{0.05}$\\
59095.07 & $15.66\pm{0.02}$ & $15.20\pm{0.04}$ & $14.71\pm{0.05}$\\
59099.08 & $15.70\pm{0.02}$ & $15.25\pm{0.02}$ & $14.75\pm{0.04}$\\
59138.93 & $15.76\pm{0.01}$ & $15.29\pm{0.03}$ & $14.83\pm{0.04}$\\
 \hline
\end{tabular}
\label{tab:IR_data}
\end{table}

The optical spectroscopic observations were performed on the 2.5-m telescope of the Caucasian Mountain Observatory of Sternberg Astronomical Institute of Moscow State University \citep{kornilov14,shatsky20} equipped with the Transient Double-beam Spectrograph (TSD\footnote{\url{http://lnfm1.sai.msu.ru/kgo/instruments/tds/}}, \citealt{potanin20}) in the 3500-7400 {\AA} range with a spectral resolution of $R\sim 1500$. The first spectrum was obtained on March 27, 2020, but most of the observations (nine nights) were carried out in August--October, 2020. During each observational night, from two to ten spectra were acquired with the exposure time of $\simeq20$ min, which provides a signal-to-noise ratio of $10-30$ for a single spectrum. The data reduction was performed in a standard way as described in \citet{potanin20}. It includes dark frame subtraction, cleaning for cosmic rays, 2D wavelength calibration with a Ne-Kr-Pb arc lamp, and flat-field correction.

The radiation flux was integrated within a three-arcsec aperture, the background was estimated using the clean regions above and below the target and removed before the integration. The flux was calibrated using ESO spectral standards\footnote{\url{https://www.eso.org/sci/observing/tools/standards.html}}. However, since we have used a narrow slit, the absolute calibration was impossible due to the slit losses. Using  the night-sky emission lines, the wavelength calibration was corrected to an accuracy of 0.2 {\AA} and 0.1 {\AA} in the blue and red beam of the TDS spectrograph, respectively. 

Additionally, photometric observations in $g,r,i$ bands were obtained from March 27 to October, 2020, with the automated 60-cm RC600 telescope of CMO SAI MSU \citep{berdnikov20}. Optical monitoring of the AT2019wey object carried out in summer-autumn 2020 revealed that AT2019wey demonstrated some variability from 0.1 magnitude (in the r-band) to 0.2 magnitude (in the g-band) during 3-4 hours of observations. In this case, the accuracy of individual measurements was from 0.01 (r-band) to 0.02 (g-band) magnitude. 

Near infrared $J,H,K$ photometry was performed from August to October 2020 with the ASTRONIRCAM camera-spectrograph of the 2.5-m telescope CMO SAI MSU \citep{nadjip17}. Photometric measurements were converted to \texttt{pha}-files using \texttt{flx2xsp} procedure from \texttt{FTOOLS} package. The measurement results are presented in the Table \ref{tab:IR_data}.

\begin{figure}
  \centerline{\includegraphics[width=0.95\columnwidth]{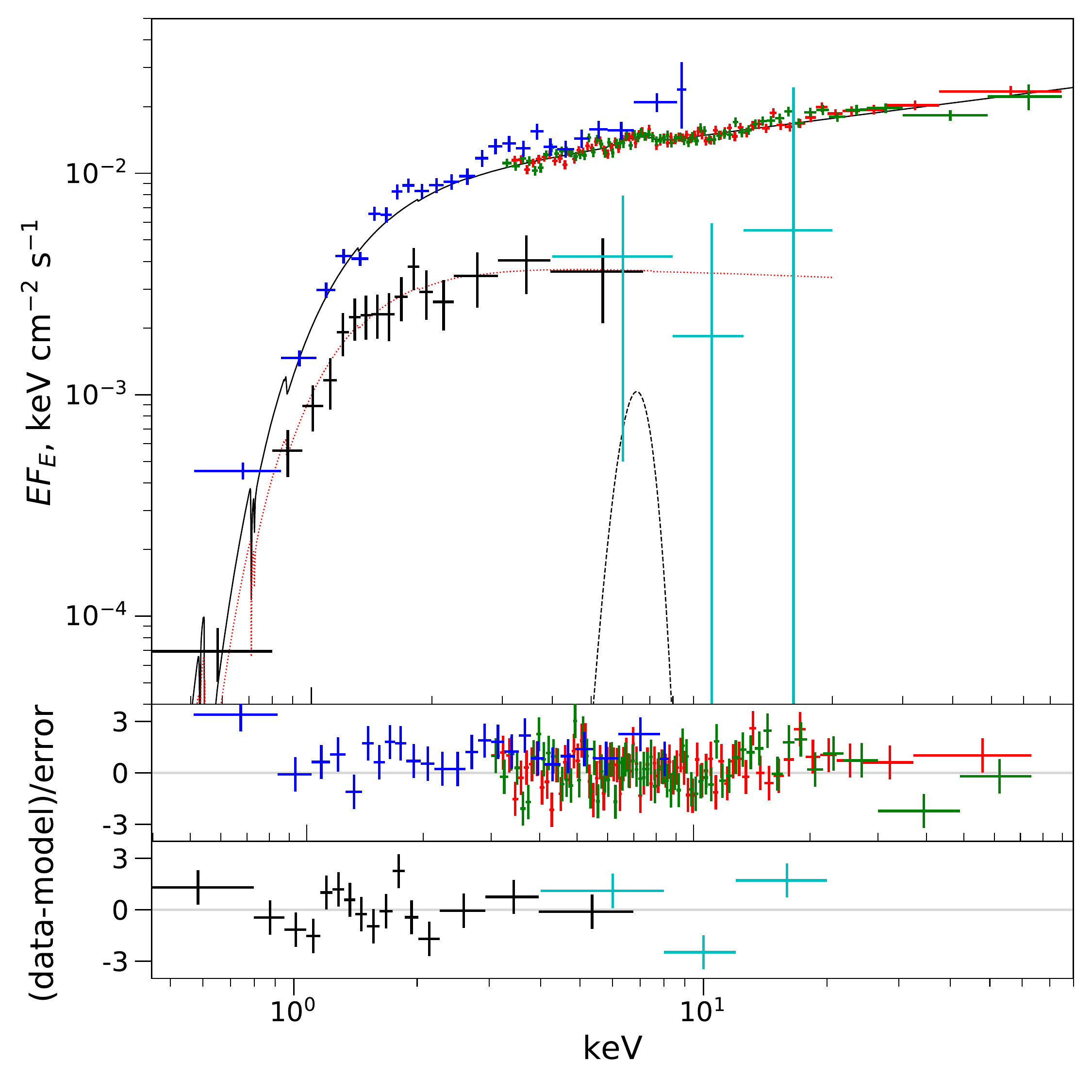}}
    \caption{X-ray spectra of \srga\ obtained in March-April, 2020. Spectrum at the discovery is shown in black (\ero) and cyan (\art) and scaled down by a factor of two, for clarity. Spectrum of the first \nust\, observation shown by green and red points (\nust\, A and B mirror modules, respectively) with the blue points corresponded to \sxrt data. The solid black line shows the best-fit model for the \nust\,+\sxrt observation, the dashed black line indicates a Fe $K_{\alpha}$ line contribution. The dashed red line corresponds to the best-fit model for the \srg\ observation.}
\label{fig:spe_spring}
\end{figure}

\subsubsection{{\it Swift/UVOT}}

The {\it Swift}/UVOT \citep{roming05} data were used to expand our SEDs into the ultraviolet range. We performed a photometry in different filters using standard preprocessed images and the \texttt{uvotsource} tool, extracting the source counts from the recommended 5$\arcsec$ radius.

The UVOT observations performed in April, 2020, were co-added to improve statistics and to secure the source detection in bluer filters. Given the lack of a strong week-scale variability in both the optical and X-ray bands, we concluded that the source's variability is negligible.

\section{Observational properties}\label{sec:obsprop}

\subsection{Long-term evolution}

A combined long-term optical -- X-ray light curve of \srga\, is shown in Fig.~\ref{fig:opt_x_lc}. Long-term optical photometric observations in $g,r$-filters from ZTF \citep{ztf19} were obtained through the $\texttt{Lasair}$ alert broker \citep{smith19_lasair} and augmented with the {\it Gaia} measurements.

The latest non-detection of the source by {\it Gaia} allows us to put an upper limit on the source brightness immediately before the outburst, assuming that during the single scan {\it Gaia} reaches down to at least 19 magnitude in the G-filter. The light curve clearly shows that after a sharp rise the source underwent a long steady decay with a mean rate of 0.04 magnitude per day in the ZTF $r$-filter. After two months from the beginning of the outburst, AT2019wey reached a plateau at the $\approx 17.65$ $r$-filter magnitude. After the gap due to the lack of the source visibility, it was found to be bright again, with another long plateau, decaying at approximately 0.01 magnitude per day. Two filters ZTF photometry suggests no obvious colour evolution throughout the outburst.

To trace variability of the X-ray emission, we produced light curves based on all available \sxrt\, and \nice\, observations. The \sxrt\, light curve in the 0.3-9 keV energy band was binned per observation, while for \nice\, we chose a bin of 1 ks in the 0.4-9 keV energy band. We then scaled the \sxrt\, measurements by a factor of 12 in order to roughly match the normalisation of \nice\, observations. A comparison of the optical (ZTF and {\it Gaia}) and X-ray long-term light curves presented in Fig.~\ref{fig:opt_x_lc} shows that despite a strong variability in X-rays, the optical light curve is nearly featureless.

\begin{table}
\centering
\caption{Best-fit parameters of X-ray spectrum, obtained with \nust / \sxrt\, in April, 2020. $F_{X}$ is estimated observed flux in the 0.1-100 keV band.}
\begin{tabular}{c | c | c } 
 \hline
 Parameter & Value & Unit\\ 
 \hline
  $N_{H}$          & $0.86^{+0.9}_{-0.9}$     & $\times10^{21}$ cm$^{-2}$\\ 
 $\Gamma$          & 1.77$^{+0.01}_{-0.01}$   & \\
 $E_{gauss}$       & 6.4 (frozen)   & keV \\
 $\sigma_{gauss}$  & 0.57$^{+0.51}_{-0.21}$ & keV\\
 $N_{phot}$        & 3.6$^{+1.8}_{-1.3}$& $\times10^{-5}$\\  
 $C_{FPMB}$        & 1.033$^{+0.015}_{-0.013}$  &\\
 $C_{XRT}$         & 0.69$^{+0.04}_{-0.04}$     &\\
$F_{X}$           & 1.11$^{+0.01}_{-0.02}$    & $\times10^{-10}$ ergs cm$^{-2}$ s$^{-1}$\\  
$\chi^{2}/d.o.f.$ & 1012.05/1049    &\\ 

  \hline
\end{tabular}

\label{tab:spring_fit}
\end{table}

\subsection{The spring 2020 observations: low-hard state}

The first X-ray spectrum of \srga\, was obtained by \srg\, on 18 March, 2020. The joint \ero/\art\, spectrum, shown in Fig. \ref{fig:spe_spring}, was described with a simple absorbed power law (\texttt{tbabs*pow} in the \texttt{XSPEC} notation). Hereafter we have used elemental abundances from \cite{wilms00} and cross-sections from \cite{verner96}. The absorbing column density was measured to be $N_{H} = 8.4^{+1.1}_{-1.0}\times10^{21}$ cm$^{-2}$ and the photon index was $\Gamma = 2.07^{+0.22}_{-0.21}$. The total observed flux in the 0.2-20 keV energy band can be estimated as $F_{X} = (3.2\pm0.5)\times10^{-11}$ erg cm$^{-2}$ s$^{-1}$. All errors correspond to a $90\%$ confidence interval. 

We then employed the first \nust\, observation, performed on 18, April, 2020, and merging first four \sxrt\, observations to study the broadband X-ray spectrum with the higher sensitivity. The X-ray spectra shown in Fig.\ref{fig:spe_spring} could be described with a simple power-law with the photon index of $\Gamma\approx$1.8 in the whole 0.5-80 keV range with an addition of a weak Fe K$_\alpha$ line. Best-fit parameters of the {\texttt const*tbabs*(pow + gauss)} model are listed in Table\,\ref{tab:spring_fit}. Parameters of the Fe K$\alpha$ line are not well constrained, and we decided to freeze its centroid at 6.4 keV during the fit. Notably, neither the spectral shape of \srga\ or its flux changed significantly over a month between the discovery and the \nust\, observation. 

\begin{figure}
  \centerline{\includegraphics[width=\columnwidth]{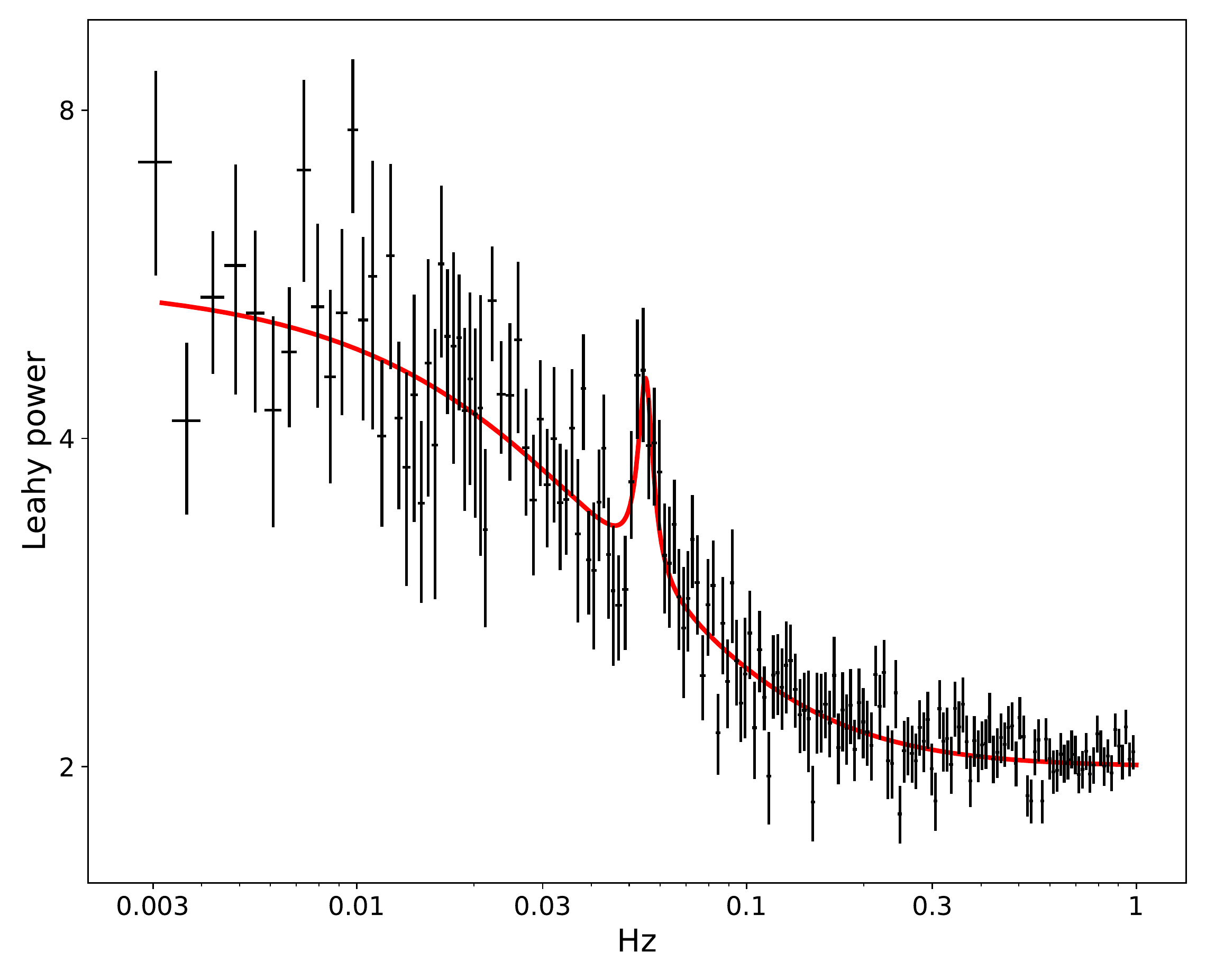}}
    \caption{The power density spectrum (in the Leahy units) of the first \nust\, observation. On top of a strong low-frequency noise, a prominent QPO peak is seen at 55 mHz.}
\label{fig:nupower}
\end{figure}

Using \nust\, data we also extracted the source light curve in the 3-78 keV energy band. After the barycentric correction, the power spectrum of the source variability exhibits the typical shape for black hole transients with a strong low-frequency noise (LFN) and a narrow peak of quasi-periodic oscillations (QPO) with a central frequency of $\simeq55$ mHz (Fig.~\ref{fig:nupower}). From the overall shape of the power spectrum we can identify the QPO to be of the type C, according to the classification scheme of \cite{casella05}. Although type-C QPOs below 0.1 Hz are rare, they were previously observed in several transient black-hole binaries, e.g., in MAXI J1820+070 \citep{buisson19}.  

The energy spectrum and timing properties suggest that during the spring of 2020 the source was in a classical 'low-hard' state (see, e.g., \citealt{homan05, gilfanov10} for the state definitions). Similar states are often observed during the rise stage of outbursts, including in short-period black-hole binaries, which are thought to be the parent class of \srga\, \citep{yao20xray}. In particular, in early observations of MAXI J1659-152 \citep{kennea11}, a similar hard spectrum was observed along with QPOs at 148 mHz, although the source was rapidly brightening during those observations, while \srga\, remained at a similar flux level for at least a month between March, 18 and April, 18. Perhaps, a more relevant example is XTE J1118+480, where a similar combination of the low-frequency QPOs (85 mHz) and hard spectrum was observed during the plateau phase of a mini-outburst \citep{frontera03}.   

\subsection{The summer-autumn 2020 plateau}

Optical observations in July-August, 2020, found the source to be still active. The {\it ZTF} light curve demonstrates a long, smooth plateau, with a gradual decay of about one millimagnitude per day in the $r$-filter. During this stage, the source was extensively monitored by \sxrt\, and \nice. The X-ray count rates presented in Fig. \ref{fig:opt_x_lc} suggest that the X-ray emission significantly varied during this period. A detailed analysis of NUII observation by \citet{yao20xray} shown that X-ray spectrum undergone drastic change, with appearance of a thermal disk emission and strong reflection component. Applying complex spectral model, consisting of a thermal disk emission up-scattered by a comptonizing electron cloud \citep{steiner17} \texttt{simplcutx(diskbb)} plus the relativistic reflection model \texttt{relxillCp} \citep{gracia14,dauser14} \citet{yao20xray} estimated physical parameters of the accretion disk - its inclination and the radius of the inner boundary. Using the data from \nice\, monitoring they also shown that the variability of X-ray flux is primarily due to an accretion disk, while the power-law component, usually associated with a coronal emission, remains relatively stable. Since this conclusion was drawn from the analysis of data below 10 keV and it is interesting to take an another look on a variability of the high-energy emission. Notable, no strong correlation between soft X-rays and optical brightness of  \srga\, was seen. 

At first, we examined two long \nust\, observations obtained at this stage (NUII and NUIII). 
We used a simple \texttt{pexrav} model \citep{pexrav} to describe only a hard X-ray part of the spectrum above 10 keV and study its evolution over time. 
For source spectra in both \nust\, observations this model has three free parameters defining the shape of the comptonized emission: power-law index $\Gamma$, reflected fraction $R$, and  cut-off energy $E_{cut}$. 
The cut-off energy was not well constrained, providing only lower limits (90\%) of $>$100 keV and $>$200 keV for NUII and NUIII, correspondingly. 
Therefore it was fixed at 1000 keV in following approximations. 
In total, in the course of $\approx$10 days, while the reflected fraction $R$ has grown from 0.31$^{+0.08}_{-0.07}$ to 0.49$^{+0.09}_{-0.08}$, the power law index softened from 1.76$^{+0.02}_{-0.02}$ to 1.91$^{+0.02}_{-0.02}$. 
We also estimated hard X-ray fluxes (in the 10-100 keV band) for both observations as 9.4$\times10^{-10}$ and 8.9$\times10^{-10}$ erg cm$^{-2}$ s$^{-1}$, respectively.

\begin{figure}
  \centerline{\includegraphics[width=\columnwidth]{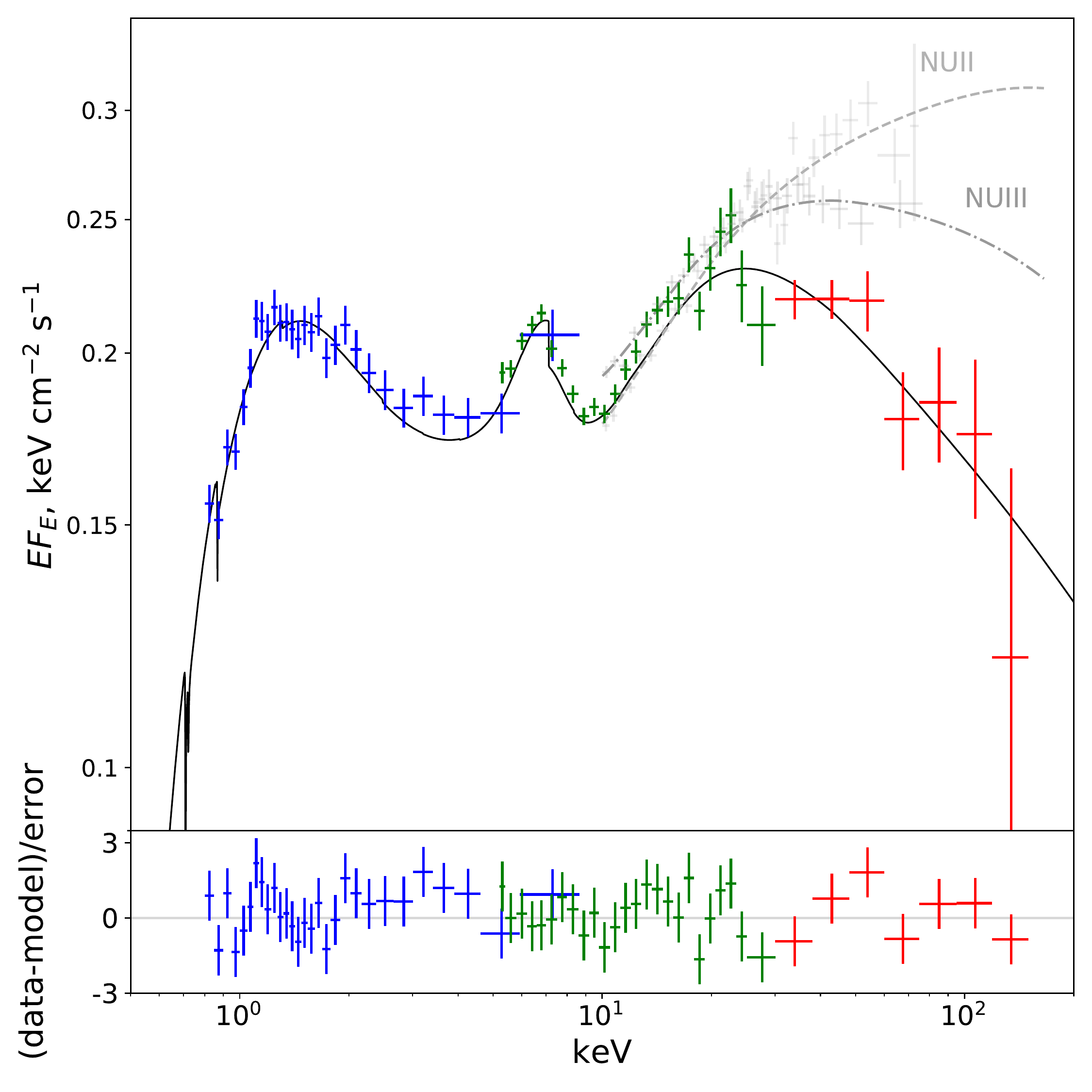}}
    \caption{Energy spectra  of the joint \sxrt - \art - {\it INTEGRAL/}IBIS observation. Blue points are from \sxrt\, green points represent \art\, data and red are {\it INTEGRAL/}IBIS measurements. Solid black line shows the best-fit model, dashed and dot-dashed lines are best fits to NUII and NUIII data, shown with grey crosses.}
\label{fig:spe_joint}
\end{figure}
\begin{table}[h]
\centering
\caption{Best-fit parameters of X-ray spectrum, obtained with \igr / \art / \sxrt\, in early October, 2020. $F_{X}$ is observed flux in the 0.1-200 keV band. }
\begin{tabular}{c | c | c } 
 \hline
 Parameter &  Value & Unit\\ 
 \hline
  $N_{H}$           & $2.4^{+0.5}_{-0.5}$   & $\times10^{21}$ cm$^{-2}$\\ 
  $kT$              & $0.36^{+0.06}_{-0.02}$   & keV \\
  $R_{in}$          & $28^{+27}_{-10}$ &  km at 10 kpc\\
  $\Gamma$          & 2.14$^{+0.09}_{-0.07}$   & \\
  $E_{cut}$         & >700    & keV \\
  $R_{refl}$        & $1.3^{+0.3}_{-0.3}$                       & \\
  $E_{gauss}$       & 6.44$^{+0.23}_{-0.29}$ & keV \\
  $\sigma_{gauss}$  & 1.01$^{+0.26}_{-0.23}$ & keV\\
  $N_{phot}$        & 1.6$^{+0.5}_{-0.4}$ & $\times10^{-3}$\\  
  $C_{IBIS}$        & 1.22$^{+0.09}_{-0.09}$  &\\
  $C_{ART-XC}$      & 0.84..0.95 &\\
$F_{X}$             & 1.77$^{+0.04}_{-0.03}$  & $\times10^{-9}$ ergs cm$^{-2}$ s$^{-1}$\\  
$\chi^{2}/d.o.f.$   & 467.43/505 &   \\ 

  \hline
\end{tabular}
\label{tab:pexrav_fit}
\end{table}

Motivated by the hardness of the observed spectrum, we initiated joint follow-up observations with the \art\, {\it INTEGRAL} and \sxrt\,  in order to probe even higher energies and to study the evolution of the high-energy part of the spectrum. In order to describe a broadband energy spectrum we extended our model, adding components that represent the multicolor blackbody disk and iron fluorescent line (i.e. \texttt{tbabs(diskbb + ga + pexrav)}). This model provides an adequate description of the observed spectrum (see Fig.~\ref{fig:spe_joint} and Table~\ref{tab:pexrav_fit}).

Interestingly, even the addition of the {\it INTEGRAL} data is not enough to securely constrain the cut-off energy $E_{cut}$, yielding again only a lower limit of $E_{cut} > 700$ keV. From the spectral fit it is also evident that the incident power-law became slightly softer, with $\Gamma=2.14^{+0.08}_{-0.08}$. The reflected fraction $R$ continued to grow, while the 10-100 keV X-ray flux decreased to 7.6$\times10^{-10}$ erg cm$^{-2}$ s$^{-1}$. Thus we can conclude, that despite of $\sim20$\% decrease in the hard X-ray flux over 50 days, no corresponding changes in the optical brightness is observed. 

Long {\it INTEGRAL} observation also provided the possibility to study a long-scale variability of \srga. We analysed source light curves obtained by IBIS, JEM-X and \art\, instruments, but no periodic variability has been found in data on the 0.1-5000 s timescales.

Overall, the \srga\ spectral shape measured during observations at the plateau stage indicates that the source was in the 'hard-intermediate' state, with the prominent and variable thermal disk and hard non-thermal emission with the strong reflection component. We found a significant spectral variability above 20 keV, but there is no corresponding variability in the optical emission (see Figs.\,\ref{fig:opt_x_lc} and \ref{fig:spe_joint}). 

\begin{figure}
  \centerline{\includegraphics[width=\columnwidth]{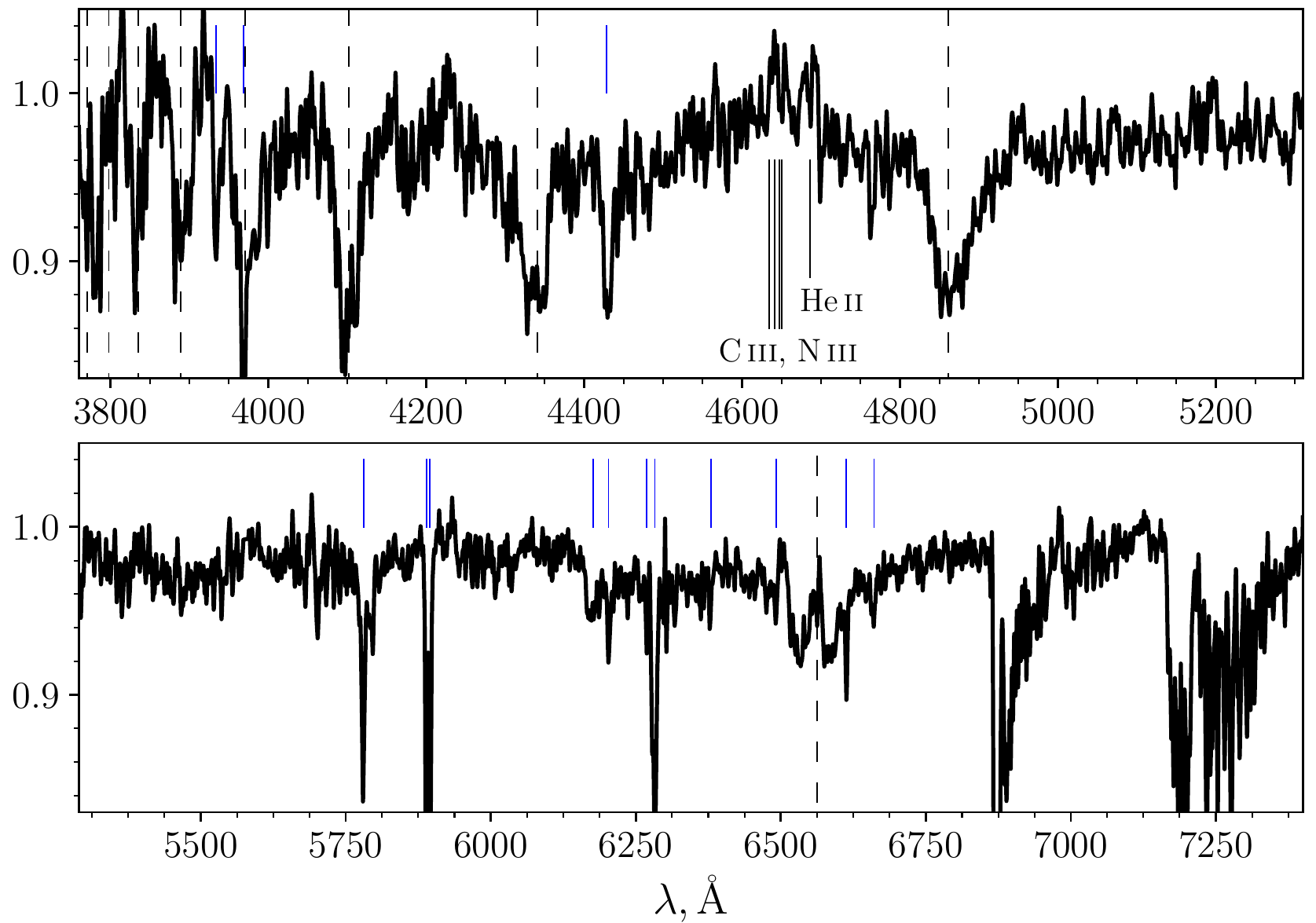}}
    \caption{The averaged optical spectrum. Positions of Balmer lines are marked with the dashed lines. Many narrow features have interstellar origin and marked with the blue lines. }
\label{fig:opt-spec}
\end{figure}

\subsection{Optical observations at CMO SAI telescopes}
\label{s:optical}

We have obtained a few dozens of optical spectra of the source with the CMO SAI 2.5-m telescope at different dates between March 2020 and November 2020. 
The spectrum exhibits a significant intra-night and night-to-night variability, mainly produced by continuum variations.

\begin{figure*}
  \centerline{\includegraphics[width=17cm]{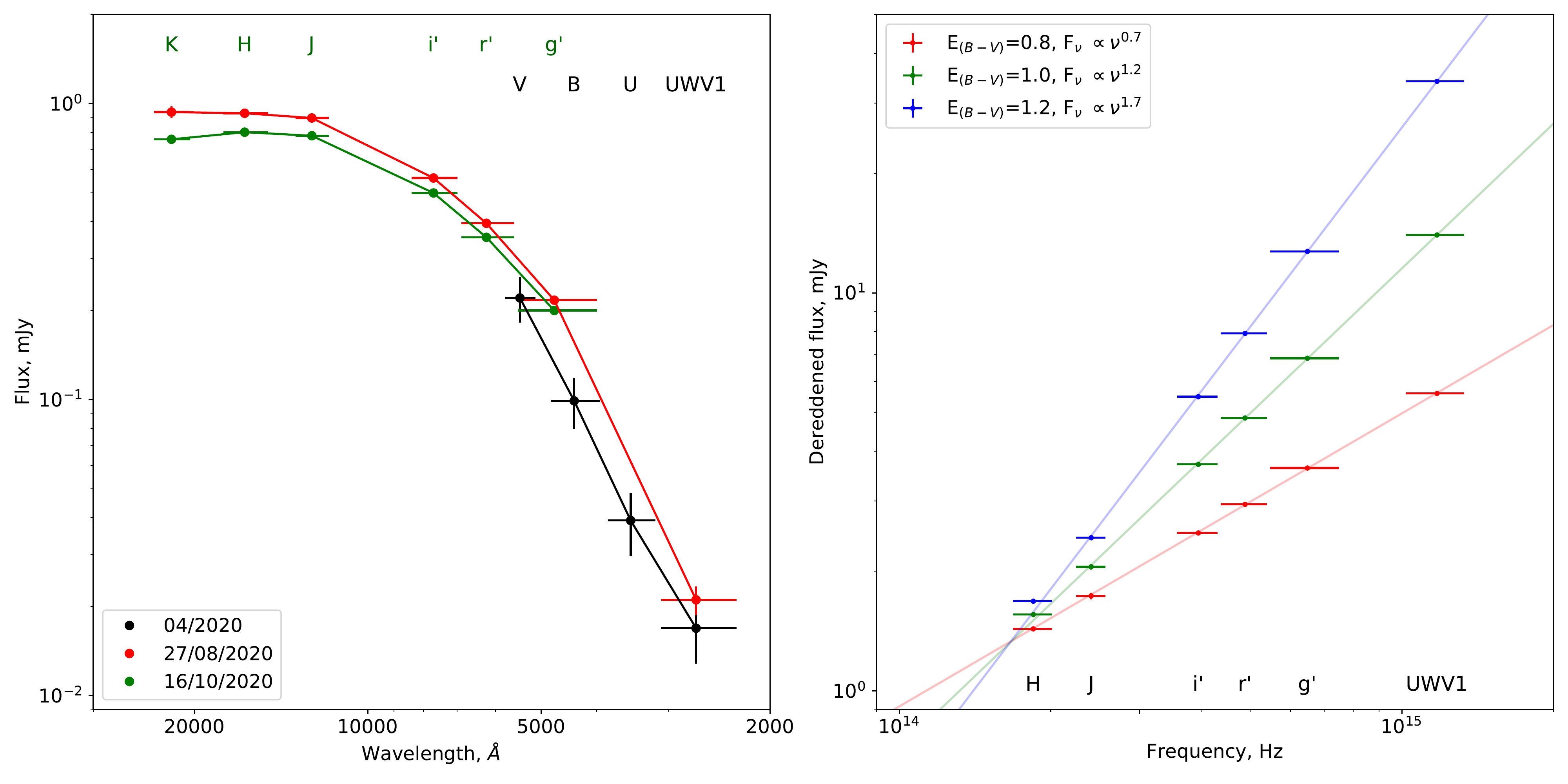}}
    \caption{{\it Left panel:}  The NIR-optical-NUV spectral energy distributions. The black points denote measurements obtained with UVOT in April, 2020; the red points mark quasi-simultaneous CMO and UVOT observations around August 27, 2020; the green point shows the CMO photometry on October 16, 2020. {\it Right panel:} The August CMO-UVOT composite SED, dereddened using different extinction coefficients (E$_{B-V}=0.8, 1.0, 1.2$) and fit with power-law models.}
\label{fig:opt-sed}
\end{figure*}

The stacked normalised optical spectrum is presented in Fig. \ref{fig:opt-spec}. The hydrogen Balmer absorption lines (H$\alpha$ - H$\epsilon$) and diffuse interstellar band features are clearly seen. The H$\alpha$ line profile is composed of a broad absorption component and a narrow emission core. Weaker emission lines are clearly seen in the spectrum at 4600-4700$\AA$, including the Bowen blend of \ion{C}{iii},\ion{N}{iii} emission lines at 4630-4660$\AA$ and the \ion{He}{ii} 4686$\AA$ line. This blend is thought to arise on the surface of the optical star experiencing strong UV-X-ray irradiation from the accretion disk and corona \citep{schachter89}. A detection of this feature gives an another possibility for the measurement of the orbital motion, which is crucial for a determination of the mass of the compact object. Although, since this blend is not readily seen in single spectra obtained with large telescopes \citep{yao20mw} such measurements will likely require a massive observational campaign.  

Using all available data we tried to trace the evolution of SED in the range from IR to NUV during the outburst. In the left panel of Fig.~\ref{fig:opt-sed} three sets of photometric measurements are shown: one obtained by {\it UVOT} in April, 2020, during the low-hard state and two sets of optical-NIR points acquired with CMO SAI and {\it UVOT} in August-October, 2020.

\citet{yao20mw} constrained the interstellar extinction towards the source to be $0.8 \lesssim E_{V} \lesssim 1.2$ using measurements of equivalent widths of several diffuse interstellar band (DIB) features. Therefore we tried to fit the composite NUV-OIR SED obtained in August with the {\texttt redden*pow} model \citep{cardelli89}, using three fixed values of reddening $E_{(B-V)} = 0.8, 1.0, 1.2$. These single power-law fits poorly the data, especially in NIR, indicating a more complex shape of the underlying continuum, which is not unexpected. In particular, a presence of a collimated jet was shown during radio observations on August 28, 2020 \citep{yao20mw}. Such a jet can contribute to the NIR continuum, while the irradiated disk and the secondary star dominates in the UV-optical emission. We, therefore, excluded a $K$-band measurement from the fits to limit the jet contribution to SED. Resulting fits are shown in the right panel of Fig.\ref{fig:opt-sed}. Note that it is hard to securely constrain the SED shape in the absence of precise measurements of the extinction towards the source.

\section{Discussion}
\label{sec:disc}
It is worth noting that the optical spectra of \srga\ display broad absorption and emission lines (see Fig. \ref{fig:opt-spec} and the discussion in \cite{yao20mw}). Model optical spectra of irradiated accretion disks can demonstrate both absorption \citep{1996AstL...22...92S} and emission 
lines \citep{2021MNRAS.tmp..821H}, depending on the X-ray illumination from the inner parts. In these models, the viscosity alpha-parameter is also significant: the smaller alpha, the smaller the equivalent width of absorption lines (mostly hydrogen Balmer lines). A strong external illumination of the disk makes the absorption lines shallow and even can invert them into emissions \citep{2021MNRAS.tmp..821H}. Therefore, accurate modelling of the disk line spectrum can probe the accretion physics.

The broadband continuum emission of accretion disks is usually described by the disk black-body model. In fact, the electron scattering should be taken into account. It is well known \citep[see, e.g.,][]{1969SvA....13..175Z,1973A&A....24..337S,1987ApJ...322..329T} that the electron scattering changes the spectrum (modified black-body radiation). Instead of the standard disk power-law dependence $f_\nu\sim \nu^{1/3}$, at high frequencies the broadband disk continuum becomes almost flat. In LMXBs with black holes, the external X-ray illumination of the outer disk also significantly modifies the low-energy part of the disk spectrum, up to inverting it to $f_\nu\sim \nu^{-1}$ (usually in the IR-bands); at even lower frequencies, the spectrum becomes the Rayleigh-Jeans continuum \citep[see, e.g.,][]{2008bhad.book.....K}.

By adopting the standard interstellar absorption (see Section \ref{s:optical} above), we can find the power-law approximation to the broad band IR-UV spectrum of the source. Depending on the assumed $E_{B-V}$, within the allowed ranges from 0.8 to 1.2 \citep{yao20mw}, the power-law continuum of the quasi-simultaneous SAI CMO - \textit{Swift}/UVOT observations (August 27, 2020; the red point in Fig. \ref{fig:opt-sed}, left panel) can be approximately fit as  $f_\nu\sim \nu^\alpha$ with $\alpha=0.7, 1.2, 1.7$  for $E_{B-V}=0.8, 1.0, 1.2$, respectively (Fig. \ref{fig:opt-sed}, right panel). This is  close to what is observed by other telescopes ($\alpha\approx 0.75$ for $E_{B-V}=0.8$  for March 2020 observations, \citealt{yao20mw}).
For $E_{B-V}\simeq 1$, these power-law approximations can be consistent with model accretion disk calculations $\alpha\approx 0.9$ \citep{1996AstL...22...92S,2021MNRAS.tmp..821H}. However, an account of the external X-ray irradiation of a modified black-body disk atmosphere would tend to flatter the continuum (i.e. decrease the model power-law photon index from 
$f_\nu\sim \nu^{0.9}$), and could explain $f_\nu\sim \nu^{0.75}$ obtained for smaller $E_{B-V}=0.8-0.9$. This issue can be quite interesting and important for black hole transients, but beyond the scope of the present study and should be further investigated. 

\section{Summary}

\label{sec:summary}
The {\it Mikhail Pavlinsky} \art\, discovery of \srga\,, the X-ray counterpart of AT2019wey,  triggered an extensive observational campaign of the source.
Follow-up observations in radio, optical and X-rays allowed to establish \srga\, as a new LMXB and a Galactic microquasar in outburst.
The detection of LF QPOs in the early stage of the outburst provides another strong indication that during Spring 2020, the source was in typical low-hard state of black-hole X-ray binaries, yet the rise time of the outburst in X-rays was significantly longer than usual for such systems. The lack of strong $L_{opt}-L_{X}$ correlation is also unexpected for such systems. Although we do find some evolution in high-energy emission of the source during the long plateau stage, it is still unclear how to interpret correctly  its broadband SED with possible contributions from the accretion disk, corona and jets. 

The detection of the Bowen blend in the stacked optical spectrum, which is likely originated in the illuminated atmosphere of the optical companion, offers the possibility of determining the system's orbital parameters. This is promising, given that \srga\, is still bright in the optical, more than a year after its discovery.

The broadband NIR-UV continuum of the source significantly changed over the period from April to October 2020. It can be roughly fitted by power-law
$f_\nu \sim \nu^\alpha$ with $\alpha\sim 0.7-1.7$, depending on the assumed interstellar reddening. The power-law of the NIR-UV continuum and shallow absorption lines could originate in an X-ray irradiated accretion disk around a compact object, presumably a black hole. The system has been in outburst since the discovery and deserves further multi-wavelength investigation.

\begin{acknowledgements}

This work is based on data from {\it Mikhail Pavlinsky} ART-XC and eROSITA, X-ray instruments aboard the SRG observatory. The SRG observatory was built by Roskosmos in the interests of the Russian Academy of Sciences represented by its Space Research Institute (IKI) in the framework of the Russian Federal Space Program, with the participation of the Deutsches Zentrum für Luft- und Raumfahrt (DLR). The eROSITA X-ray telescope was built by a consortium of German Institutes led by MPE, and supported by DLR.  The SRG spacecraft was designed, built, launched and is operated by the Lavochkin Association and its subcontractors. The science data are downlinked via the Deep Space Network Antennae in Bear Lakes, Ussurijsk, and Baykonur, funded by Roskosmos. The eROSITA data used in this work were processed using the eSASS software system developed by the German eROSITA consortium and proprietary data reduction and analysis software developed by the Russian eROSITA Consortium.
The \art\ and \ero\ teams thank the Russian Space Agency, Russian Academy of Sciences and State Corporation Rosatom for the support of the \srg\ project and \art\ telescope and the Lavochkin Association (NPOL) with partners for the creation and operation of the \srg\ spacecraft (Navigator). In this work we used also data from NuSTAR and INTEGRAL observatories. INTEGRAL is an ESA project with instruments and science data centre funded by ESA member states (especially the PI countries: Denmark, France, Germany, Italy, Switzerland, Spain) and with the participation of Russia and the USA.  This research has made use of data, software and/or web tools obtained from the High Energy Astrophysics Science Archive Research Center (HEASARC). This work made use of data supplied by the UK Swift Science Data Centre at the University of Leicester. Observations in X-rays and corresponding data analysis were performed under support of the Russian Science Foundation grant 19-12-00423, to which IAM, AAL, SAN and SVM are grateful. The work of AVD, AMT was partially supported by the RSF grant 17-12-01241 (optical and IR data reduction). 
\end{acknowledgements} 

\bibliographystyle{aa} 
\bibliography{biblio.bib}

\end{document}